\documentclass[ ]{revtex4}
\usepackage{hyperref}
\usepackage{amsmath}

\usepackage{amsfonts}

\begin{document}
\title{\bf \Large   A mass-deformed BLG theory  in Gribov-Zwanziger framework}

 \author { Sudhaker Upadhyay }
 \email{sudhakerupadhyay@gmail.com}
\affiliation { Centre for Theoretical Studies,\\
Indian Institute of Technology Kharagpur,  Kharagpur-721302, WB, India}

\begin{abstract} 
 In this paper, we extend the Gribov-Zwanziger framework accounting for the existence of Gribov copies   to the mass deformed Bagger--Lambert--Gustavsson (BLG) theory in $\mathcal{N} =1 $ superspace.
The restriction
of the domain of integration in the Euclidean functional integral to the first Gribov horizon is
implemented, by adding a non-local horizon  term to the effective action. 
Furthermore, the soft breaking of the BRST symmetry  due to horizon term is restored with the 
help of external sources. We compute the various Ward identities for this theory
relying on the Lie 3-algebras.
\end{abstract}

\maketitle
 
\section{Introduction}
The understanding of M-theory is important from the view point that it can be the most
profound  unified theory for particle physics. So,  it is
essential to study the behavior of M-branes, a basic ingredient of M-theory.
 Dirac's prescription on monopoles suggests that charge of M-brane is quantized, i.e.,
 the number of charges are countable. 
While the
dynamics of a single M-brane is well understood,  at least at classical level, a very little
is known about the multiple M-branes.  
In such circumstances,    a lot of excitements have been seen for a model of
multiple M2-branes  based on Lie 3-algebra proposed by Bagger, Lambert \cite{bl,b2,b3} and Gustavsson \cite{g}. The Bagger--Lambert--Gustavsson  (BLG) model describes a
 three dimensional
superconformal field theory with ${\cal N} = 8$ supersymmetry, proposed as
the world-volume action for two interacting M2-branes. In this model, the field content
is a collection of scalars, fermions and gauge fields transforming under a   Lie 3-algebra (a generalization of a Lie algebra with a triple bracket replacing the commutator and a
4-index structure constant replacing the usual 3-index structure constant of a Lie algebra).

Although the original BLG action  possess  high amount of supersymmetry \cite{ben,ben2,ben3,ben4},
  the prominence of   simple (or ${\cal N} = 1$) superfields in three dimensions can never be  underrated \cite{ket}. The superfield description of three dimensional BLG  theory in 
${\cal N} = 1$ superspace is described in Ref. \cite{ma}.
The dimensional  reduction of the multiple M2-branes in
${\cal N} = 1$ superspace formalism has   been analysed \cite{8}. In this context, a map to a 
Green-Schwarz string wrapping a nontrivial circle in $\mathbb{C}^4/\mathbb{Z}_k$
has also been constructed. Moreover, it has   been observed that  the mass deformation breaks the conformal invariance of the   BLG theory though it
 preserves maximal supersymmetry \cite{kim,kim1}. While this deformed theory has
discrete vacua, its non-relativistic limit in the symmetric phase turns out to
 acquire a superconformal symmetry, different from the original mass-deformed ABJM model.
 For a given number
of particles and antiparticles in mass-deformed scenario, we can consider the low 
energy physics where the speed of particles are much slower than that of the speed of light.
 As a consequence of the conformal breaking, there will exist a renormalization group invariant scale, it
turns out that it is  possible to attach a dynamical meaning to the
Gribov parameter. This implies
that, in
mass deformed BLG theory,   the restriction of the domain of integration in the
functional integral to the Gribov region is important, and  non-perturbative dynamical scale can be generated.  
 Recently, the BLG theory in ${\cal N} = 1$ superspace  has been quantized in Lorentz gauge \cite{sud,sud1,sud11,sud22,sud23,sud231,sud24,sud241,sud242}. Here, we study the
 BRST quantization of mass deformed   BLG theory in ${\cal N} = 1$ superspace in Landau gauge. 
 
According to standard quantization methods of  a gauge theories, for almost all calculations aside from lattice simulations of gauge-invariant quantities, one needs to fix  a gauge \cite{1}. In certain choices, even after fixing the gauge, the redundancies of gauge fields  do  not disappear completely 
 for large value of gauge fields (Gribov problem) \cite{2}. The
non-Abelian gauge theories in those gauges contain so-called Gribov copies, which
 play a crucial role in the infrared (IR) regime while it can be neglected in
the perturbative ultraviolet regime \cite{2,3,4}. Such investigations have become very exciting currently due to the fact that color confinement is closely related to the asymptotic
behavior of the ghost and gluon propagators in deep IR regime \cite{5}. 
In order to make the theory free from those copies,   Zwanziger   proposed a theory, 
commonly known 
as  Gribov and Zwanziger (GZ) theory, by restricting
the domain of integration in the functional integral with a (non-local) horizon term
\cite{3}.  A composite fields approach has also been presented to eliminate Gribov copies from non-Abelian theories \cite{Re0}.

 On the other hand,   the horizon term of the GZ action breaks the BRST symmetry of the theory softly 
\cite{sore1}. However, the Kugo-Ojima (KO) criterion for color confinement \cite{ko} is based on the assumption of   
  BRST exactness of the theory.  It has been shown that a consistent quantization
of gauge theories with a soft breaking of BRST symmetry does not exist \cite{lav}
and leads to inconsistency in the conventional BV formalism \cite{lav1,Re}. 
Further,  It has been established that the gauge theories with soft breaking of BRST symmetry can be made consistent if the transformed BRST-breaking terms satisfy certain requirement
\cite{Re1}.
The BRST symmetry in presence of the Gribov horizon has great applicability in order to 
solve the non-perturbative features of confining Yang-Mills (YM) theories \cite{dud, fuj},
where the soft breaking of the BRST symmetry exhibited by the GZ action can be converted 
into an exact invariance \cite{sor1,sud3}. Also, the spontaneously broken Slavnov-Taylor 
identity has been converted to the linearly broken Ward identity under certain algebraic circumstances \cite{sorella}.
Subsequently, a nilpotent BRST transformation which leaves the GZ
action invariant has been obtained and  
can be applied to KO analysis of the GZ theory \cite{sor}.
Finite BRST-antiBRST transformations were also developed in the GZ context \cite{sud3,Re3}.
A full resolution of the Soft breaking of BRST symmetry was done 
in Ref. \cite{Re4}.  The GZ treatment in $R_\xi$  gauges  was done for the standard model also. Recently,
a gauge-invariant formulation of the GZ model for YM theory with local BRST transformations was    given  for the first time in Ref. \cite{Re5}.
 Such investigations are very useful in order to
evaluate the vacuum expectation value   of BRST exact quantity.

 In this work, we consider the mass deformed BLG theory $\mathcal{N} =1 $ superspace in Landau gauge.
 We derive the  BRST symmetry for the theory. Furthermore, to discuss the non-perturbative
regime,  we implement the
 GZ framework to the theory  by adding a non-local horizon term to the effective action
 which restricts the domain of integration of functional integral to the first Gribov horizon.
We also localize the horizon term by extending the configuration space 
 with quartet of auxiliary fields.  Within formulation, we notice that the presence of $\gamma$-dependent terms break the BRST invariance of the BLG action. To restore the BRST invariance of the
 BLG theory in GZ framework, we introduce   three more pairs of external sources with certain  physical  values.
This symmetry turns out to be useful in order to establish non-perturbative Ward identities, allowing
us to evaluate the vacuum expectation value of quantities which are BRST exact.
 Further, we compute the Ward identities corresponding to   BRST exact action together
 with external sources. The present investigations will be helpful to 
 compute the counter terms for the multiplicative renormalizability of the theory.

 The paper is organized as follows. In Sec. II, we analyse the mass deformed  BLG Theory in $\mathcal{N} =1 $ superspace in Landau gauge. This theory follows Lie 3-algebras.
 In Sec. III, we discuss the theory in GZ framework. Within this framework the BRST 
 symmetry of the GZ action is re-established. In sec. IV, we derive the various Ward identities 
 useful in the proof of renormalizability of the theory. In Sec. V, we conclude the results and make future remarks. 
\section{The BLG Theory in $\mathcal{N}=1$ superspace: short review}
In this section, first of all we review  the construction of BLG theory 
in $\mathcal{N} =1 $ superspace. To write the action, we first introduce 4-index structure constants $f^{ABCD}$ associated
with   a trilinear antisymmetric
product \cite{bl},
\begin{equation}
  [T^A,T^B,T^C]=f^{ABC}_{D}T^D,
\end{equation}
where $T^A$'s are the generators of this Lie 3-algebra.
A generalization of the trace,  taken over the three-algebra indices,  provides
an appropriate `3-algebra metric':
$h^{AB}= \mbox{Tr}(T^A T^B)$, 
 which  may  raise  and lower the indices.
Totally anti-symmetric in nature, i.e. $f^{ABCD} = f^{[ABCD]}$, these  structure constants  satisfy  
 the fundamental (Jacobi) identity, 
\begin{eqnarray}
f^{[ABC}{}_G f^{D]EG}{}_H =f^{AEF}{}_G f^{BCDG}- f^{BEF}{}_G f^{ACDG}+f^{CEF}{}_G f^{ABDG}-f^{DEF}{}_G f^{ABCG}= 0.  
\end{eqnarray}
Another quantity comprised with 4-index structure constants, $C^{AB,CD}_{EF} = f^{AB[C}_{[E} \delta^{D]}_{F]}$,
are antisymmetric in the pair of indices $AB$ and $CD$ and 
satisfy  \cite{bl00}
\begin{equation}
C^{AB,CD}_{EF} C^{GH,EF}_{KL} + C^{GH,AB}_{EF} C^{CD,EF}_{KL} 
+C^{CD,GH}_{EF} C^{AB,EF}_{KL} = 0.
\end{equation}

In order to construct BLG theory in $\mathcal{N} =1 $  superspace, we first define the
non-Abelian    Chern-Simons  action   as 
\begin{equation}
S_{CS} =-\frac{k}{4\pi} \int d^3x \, \nabla^2 
 [ f^{ABCD}  \Gamma^{a}_{AB}    \Omega_{a CD}]_|,
\end{equation}
where $k$ is an integer and 
\begin{eqnarray}
 \Omega_{AB a} & = & \omega_{a AB} - 
\frac{1}{6}C^{CD, EF}_{AB} \Gamma^b_{CD} \Gamma_{ab EF} \\
 \omega_{AB a} & = & \frac{1}{2} D^b D_a \Gamma_{AB b} 
- \frac{i}{2}  C^{CD, EF}_{AB}\Gamma^b_{CD}  D_b \Gamma_{a EF} 
  -
 \frac{1}{6} C^{CD, EF}_{AB} C_{EF}^{LM, NP} \Gamma^b_{CD} 
 \Gamma_{b LM } \Gamma_{a NP}, \label{omega} \\
 \Gamma_{AB ab} & = & -\frac{i}{2}  \left[D_{(a}\Gamma_{AB b)} 
- iC^{CD, EF}_{AB}\Gamma_{a CD}  \Gamma_{b EF} \right].
\end{eqnarray}
with the super-derivative  defined by
$
 D_a = \partial_a + (\gamma^\mu \partial_\mu)^b_a \theta_b
$.

The matter action    is given by 
\begin{eqnarray}
S_{M} &=& -\frac{1}{4} \int d^3x  \, \nabla^2 
  \left[ \nabla^a_{}          X_A^{I \dagger}          
\nabla_{a }          X^A_I +m^2 X_A^{I \dagger}X^A_I +\mathcal{V}_{   } \right]_|,
\end{eqnarray}
where $m$ refers to the mass, which breaks the conformal invariance,  and  the covariant derivative  is given by 
\begin{eqnarray}
\nabla_{a}         X^{ A I } = 
D_a  X^{ A I } + i \Gamma^{AB}_{a  }         X^{I }_B.
\end{eqnarray}
The potential term $ \mathcal{V}$ is defined by 
$
 \mathcal{V}  = \frac{8\pi}{k} f_{ABCD}\epsilon^{IJKL} 
 [ X^A_I  X^{B\dagger}_K  X^C_J  Y_L^{D\dagger}] 
$.

Now,  the classical BLG action in $\mathcal{N} =1 $  superspace is given by \cite{af}
\begin{eqnarray}
{ S_c} =S_{CS} +   S_{M}.\label{act1}
\end{eqnarray}
The fields  of the above action transform 
under the gauge  transformation  as follows,   
\begin{eqnarray}
\delta X^{IA }= i(\Lambda{     } X^{I })^A,\ \   
  \delta  X^{IA \dagger  } 
= -i(X^{IA\dagger  }{     } \Lambda)^A, \ \
  \delta \Gamma_a^{AB} =(\nabla_a {     } \Lambda)^{AB}.  \label{g}
\end{eqnarray}
The BLG action   (\ref{act1}) remains invariant under these  gauge transformations.
This implies that there are redundancy in the gauge degrees of freedom  of  
 the  BLG  action and thus all gauge degrees of freedom are not physical. 
 To get rid of  such spurious degrees of freedom,  we need to fix a gauge before performing
  any 
calculations. Here, we  fix the gauge with a  suitable  covariant  gauge  condition, 
$
G =   D^a  \Gamma_a ^{AB} =0
$.
This  gauge fixing condition  can be
incorporated in the action at the quantum level by adding the following   Landau
gauge fixing term to 
the classical action (\ref{act1}),
\begin{equation}
S_{gf} = \int d^3x  \, \nabla^2
 \left[f^{ABCD}b_{AB}  D^a \Gamma_{a CD}   
\right]_|.
\end{equation}
The induced Faddeev-Popov ghost term,  corresponding to the above gauge fixing term,  is given by
\begin{equation}
S_{gh} = \int d^3x  \, \nabla^2
\left[ f^{ABCD}\bar{c}_{AB}   D^a \nabla_a   c_{CD}
\right]_|.
\end{equation}
 Now,    the effective action for 
  BLG theory in Landau gauge in ${\cal N}=1$ superspace  is given as the sum of the classical action to the gauge 
fixing term and the ghost term,
\begin{equation}
 S_{BLG} =S_c +S_{gf}+  S_{gh}.\label{action}
\end{equation}
This effective action  enjoys the following nilpotent BRST symmetry:
\begin{eqnarray}
s \,\Gamma^{AB}_{a} = -[\nabla_a   c]^{AB}  , 
&&
s \,c_{AB} =  \frac{1}{2}C^{CD, EF}_{AB}{c_{CD} c_{EF}}   ,
\nonumber \\
s \,\bar{c}^{AB} = b^{AB}   , 
&&
s \,b^{AB} =0,  \nonumber \\ 
s \, X^{I A } = ic^{AB}  X^{I }_{B}   , 
 &&  
s \, X^{ I A\dagger }
 =  - i  X^{I \dagger }_B c^{AB}.
\end{eqnarray}
 The sum of gauge fixing   and 
the ghost terms is BRST exact, so, it  can be expressed in terms of BRST variation,
\begin{eqnarray}
S_{gf} + S_{gh} 
 &=& s \int d^3x   \, \nabla^2\left[ f^{ABCD} \bar{c}_{AB} 
 D^a  \Gamma_{aCD} 
\right]_|.
\end{eqnarray}
In fact,  due to the nilpotency of the BRST transformations,  the invariance of the 
effective action $S_{BLG}$ under BRST symmetry is evident.

 \section{The BLG theory in Gribov-Zwanziger framework}
 In this section, we discuss the BLG theory in the GZ framework. The motivation 
 for such study is to handle covariant gauge fixing correctly as they are not ideal in
   non-perturbative (IR) regime. 
 Since   two equivalent  superfields, satisfying the   
 the Landau gauge,  connected by a gauge transformation (\ref{g}),  yield 
 \begin{eqnarray}
 \nabla^2 D^a\nabla_a\Lambda^{AB} =0.
 \end{eqnarray}
Therefore, the existence of infinitesimal copies, even after Faddeev-Popov  quantization, is related to 
the presence of the zero modes of the operator $\nabla^2 D^a\nabla_a\Lambda^{AB}$.
To see the zero mode problem, we take the eigenvalues  equation
\begin{eqnarray}
 \nabla^2 D^a\nabla_a\Lambda^{AB} =\lambda \Lambda^{AB}.
\end{eqnarray}
For configurations very close to the vacuum (i.e $\Gamma_a^{AB} =0$), this reduces to
\begin{equation}
(D^2)^2 \Lambda^{AB} =-\partial^2 \Lambda^{AB} =\lambda\Lambda^{AB},
\end{equation}
 and, therefore, the operator has only positive eigenvalues. However,
  this can not be guaranteed always and may be
displayed negative eigenvalues  at larger amplitudes than the vacuum, i.e., sufficiently large 
$\Gamma_a^{AB}$. Hence, we
analytically implement the restriction to the Gribov region $\Omega$, defined as the set of field 
configurations fulfilling the Landau
gauge condition, for which the Faddeev-Popov operator ($ - f^{ABCD}\nabla^2D^a 
\nabla_a(\Gamma_a^{AB})$) is strictly positive, as
\begin{eqnarray}
\Omega :=\left\{\Gamma_a^{AB}|  D^a  \Gamma_a ^{AB}=0,\ - f^{ABCD}\nabla^2D^a \nabla_a(\Gamma_a^{AB}) 
>0 \right\}.
\end{eqnarray} 
 The restriction to the domain of integration, in the path integral, can be achieved by
following the GZ approach,   where  the inverse of this operator  (horizon function) is included in
the functional integral in order to compensate the problem. This is achieved by 
adding the following horizon term  to the effective BLG action:
\begin{eqnarray}
S_h=\int d^3x\ h(x)=\gamma^4\int d^3xd^3y \  \nabla^2 \left[ C_{AB}^{CD, EF} \Gamma_{aCD}(x) (-f^{ABGH}\nabla^2D^a \nabla_a)^{-1}C_{GH,EF}^{LM}\Gamma^a_{LM}(y)\right]_|,
 \end{eqnarray}
 and the resulting action is being  called as the GZ
BLG action. Here, the parameter, $\gamma$, has the dimension of  the mass and is known
as the Gribov parameter. This is not a free parameter. This is a dynamical quantity, being
determined in a self-consistent way through a gap equation, called the horizon condition,
\begin{eqnarray}
\langle h\rangle =3\gamma^4 f(N),\label{h}
\end{eqnarray}
where $f(N)$ is a some constant number. 
 
 The partition function for GZ BLG action is defined by
 \begin{equation}
 Z_{BLG}^{GZ} = \int_{\Omega} {\cal D} \Gamma{\cal D}X
 {\cal D}X^\dag{\cal D}c{\cal D}\bar c{\cal D} b \ e^{-S_{BLG}} = \int{\cal D} \Gamma{\cal D}X
 {\cal D}X^\dag{\cal D}c{\cal D}\bar c{\cal D} b \ e^{-(S_{BLG}+S_h-3\gamma^4 f(N))}.
 \end{equation}
  By localizing the non-local term it can further be rewritten as
  \begin{equation}
 Z_{BLG}^{GZ} = \int_{\Omega} {\cal D} \Gamma{\cal D}X
 {\cal D}X^\dag{\cal D}c{\cal D}\bar c{\cal D} b{\cal D}\bar\varphi{\cal D} \varphi{\cal D}\bar{\omega}{\cal D}\omega \ e^{-S^{GZ}_{BLG}},
 \end{equation}
 where the GZ BLG action is given by
 \begin{eqnarray}
 S^{GZ}_{BLG} =S_{BLG}+ S_0+ S_\gamma,\label{a}
 \end{eqnarray}
 with the (localized) horizon action
 \begin{eqnarray}
 S_0&=&\int d^3x  \  \nabla^2 \left[ \varphi_{aEF}^{AB}( f^{ABCD}\nabla^2 D^a\nabla_a)\varphi^{aCDEF} -\omega_{aEF}^{AB}(  f^{ABCD}\nabla^2 D^a\nabla_a)\omega^{aCDEF} \right.\nonumber\\
 &-&\left. C_{AB, CD}^{GH}(D_a\bar{\omega}_b^{ABLM})(f^{CDEF}\nabla^a c_{EF})\varphi^b_{GHLM} \right]_|
 \end{eqnarray}
 and
 \begin{eqnarray}
 S_\gamma =-\gamma^2 \int d^3x \ \nabla^2 \left[ C^{AB}_{CD,EF}\Gamma_{aAB} (\varphi^{aCDEF}
 +\bar\varphi^{aCDEF})\right]_|-3\gamma^4 f(N).
 \end{eqnarray}
In this local formulation, the horizon condition (\ref{h})  takes the following form:
\begin{eqnarray}
\frac{\partial {\cal E} }{\partial\gamma^2} =0,
\end{eqnarray}
where the vacuum energy ${\cal E}$ is defined by
$
e^{-{\cal E}}=Z_{BLG}^{GZ}$.
Here,  in the absence of $\gamma$-dependent term,  the action (\ref{a})  enjoys the 
following BRST symmetry:
\begin{eqnarray}
s \,\Gamma^{AB}_{a} = -[\nabla_a   c]^{AB}   , 
&&
s \,c_{AB} =  \frac{1}{2}C^{CD, EF}_{AB}{c_{CD} c_{EF}},
\nonumber \\
s \,\bar{c}^{AB} = b^{AB} , 
&&
s \,b^{AB} =0,  \nonumber \\ 
s \, X^{I A } = ic^{AB}  X^{I }_{B} , 
 &&  
s \, X^{ I A\dagger }
 =  - i  X^{I \dagger }_B c^{AB} ,\nonumber\\
 s \, \bar\omega_{aEF}^{AB}=  \bar\varphi_{aEF}^{AB} , 
 &&  
s \,\varphi_{aEF}^{AB}=  \omega_{aEF}^{AB}. \label{BRS}
\end{eqnarray}
The GZ BLG action is, however, not   invariant under the above set of BRST transformations, due to the term $S_\gamma$, as
\begin{eqnarray}
s S^{GZ}_{BLG}=s S_\gamma =\gamma^2\int d^3x\ \nabla^2\left[ C_{AB,CD}^{EF} f^{ABLM}\nabla_a c_{LM}(
\varphi^{aCD}_{EF}+\bar{\varphi}^{aCD}_{EF}) -C_{AB}^{CD,EF}\Gamma_a^{AB}
\omega^a_{CDEF} \right]_|.\label{16}
\end{eqnarray}
Utilizing the BRST variation, we rewrite the GZ BLG action by
\begin{eqnarray}
 S^{GZ}_{BLG} =S_c+s\int d^3x \ \nabla^2\left[ f^{ABCD} \bar{c}_{AB} 
 D^a  \Gamma_{aCD} +\bar{\omega}^{ABEF}( f^{ABCD}D^a\nabla_a)\varphi^b_{CDEF}
\right]_|+S_\gamma,
\end{eqnarray}
from which relation (\ref{16}) becomes apparent.

We construct a BRST invariant action, corresponding to the BRST breaking term $S_\gamma$,
as follows
\begin{eqnarray}
\Sigma_\gamma &=& s\int d^3x\ \nabla^2 \left[  -U_{ab}^{ABCD}f_{ABEF}\nabla^a \varphi_{CD}^{bEF}
-V_{ab}^{ABCD} f_{ABCD} \nabla_a\bar{\omega}^{bEF}_{CD}
-U_{ab}^{ABCD}V^{ab}_{ABCD} \right.\nonumber\\
&+&\left. T^{ABCD}_{ab} C^{EF}_{AB,LM} f^{LMNO}\nabla^a C_{NO} \bar{\omega}^b_{EFCO}\right]_|,
\end{eqnarray}
where we have introduced 3 new doublets $(U_{ab}^{ABCD}, M_{ab}^{ABCD}), (V_{ab}^{ABCD}, N_{ab}^{ABCD})$ and $(T_{ab}^{ABCD},R_{ab}^{ABCD})$ with the following BRST transformations:
\begin{eqnarray}
&&s\ U_{ab}^{ABCD}= M_{ab}^{ABCD},\ \ \ \ s\ M_{ab}^{ABCD}=0,\nonumber\\
&&s\ V_{ab}^{ABCD}= N_{ab}^{ABCD},\ \ \ \ s\ N_{ab}^{ABCD}=0,\nonumber\\
&&s\ T_{ab}^{ABCD}= R_{ab}^{ABCD},\ \ \ \ s\ R_{ab}^{ABCD}=0.\label{BRS1}
\end{eqnarray}
In order to make this extended theory reminiscent with the original one, 
we, therefore,  
 set  (at the end) the sources to have such values    that $\Sigma_\gamma|_{phy}=S_\gamma$.
We have, thus, restored the broken BRST symmetry, which may be helpful  to establish the
renormalizability of the GZ BLG theory.

Thus, the final BLG action in GZ framework is given by
\begin{eqnarray}
\Sigma_{BLG}=S_{BLG}+S_0+\Sigma_\gamma,
\end{eqnarray}
which remains invariant under the BRST transformations given in (\ref{BRS}) and (\ref{BRS1}).
It is apparent that the spontaneous breaking of the BRST symmetry is entirely driven by the Gribov 
parameter. This implies that the breaking is due to  the Gribov horizon, which assures that the analysis is truly non-perturbative.
\section{The Ward identities}
Now, we should try to find all the possible Ward identities. In order  to write the 
Slavnov-Taylor identity, we first have to couple all nonlinear BRST transformations to the external
sources. We find that  $\Gamma_a^{AB}, c_{AB}, X^I_A$  and $X_A^{I\dag}$ transform nonlinearly under the BRST transformation. Therefore, we add the following
term to the action $\Sigma_{BLG}$:
\begin{eqnarray}
S_{ext}=\int d^3x\ \nabla^2\left[ -K_a^{AB}(\nabla^a c)_{AB}+\frac{1}{2}L^{AB}C_{AB,CD}^{EF}c^{CD}
c_{EF}+ i \bar Y_{AI} c^{AB}X^I_B- iX_B^{I\dag}c^{AB}Y_{AI}\right]_|,
\end{eqnarray}
where
 $K_a^{AB}, L^{AB}, Y_{AI} $
 and $\bar Y_{AI}$ are four new
sources, invariant under the BRST symmetry $s$ and with the physicality conditions
\begin{eqnarray}
K_a^{AB}|_{phys}= L^{AB}|_{phys}=Y_{AI}|_{phys}=\bar Y_{AI}|_{phys}=0.
\end{eqnarray}
The enlarged action is, thus, given by
\begin{eqnarray}
\Sigma_{BLG}'=\Sigma_{BLG}+S_{ext},
\end{eqnarray}
which is indeed BRST invariant.
This action $\Sigma_{BLG}'$ now enjoys a larger number of Ward identities mentioned  below:
\begin{itemize}
\item   The Slavnov-Taylor identity is given by
\begin{eqnarray}
{\cal S}(\Sigma_{BLG}')=0,
\end{eqnarray}
where
\begin{eqnarray}
{\cal S}(\Sigma_{BLG}')&=&\int d^3x \left( \frac{\delta\Sigma_{BLG}'}{\delta K_a^{AB}}
\frac{\delta\Sigma_{BLG}'}{\delta \Gamma^a_{AB}} + \frac{\delta\Sigma_{BLG}'}{\delta L^{AB}}
\frac{\delta\Sigma_{BLG}'}{\delta c_{AB}} +\nabla^2 b^{AB} \frac{\delta\Sigma_{BLG}'}{\delta\bar c^{AB}}  
+\nabla^2\varphi^{AB}_{aCD} \frac{\delta\Sigma_{BLG}'}{\delta\bar \omega_{aCD}^{AB}} \right.\nonumber\\
&+&\left. \nabla^2\omega^{AB}_{aCD} \frac{\delta\Sigma_{BLG}'}{\delta\bar \varphi_{aCD}^{AB}} 
+\nabla^2 M^{ABCD}_{ab} \frac{\delta\Sigma_{BLG}'}{\delta\bar U_{ab}^{ABCD}} +\nabla^2 N^{ABCD}_{ab} \frac{\delta\Sigma_{BLG}'}{\delta\bar V_{ab}^{ABCD}}\right.\nonumber\\
&+&\left.  \nabla^2 R^{ABCD}_{ab} \frac{\delta\Sigma_{BLG}'}{\delta\bar T_{ab}^{ABCD}} +
\frac{\delta\Sigma_{BLG}'}{\delta\bar Y^{AI} }
\frac{\delta\Sigma_{BLG}'}{\delta X_{AI}}-\frac{\delta\Sigma_{BLG}'}{\delta  Y^{AI} }
\frac{\delta\Sigma_{BLG}'}{\delta X^\dag_{AI}}
\right).
\end{eqnarray}
\item  The $U(f)$ invariance reads
\begin{eqnarray}
{\cal U}_{ab}^{CDEF}\Sigma_{BLG}' =0,
\end{eqnarray}
with
\begin{eqnarray}
{\cal U}_{ab}^{CDEF} &=&\int d^3x \left[\varphi_{a}^{ABCD}\frac{\delta}{\delta\varphi_{b}^{ABEF}}- \bar\varphi_{b}^{ABEF}\frac{\delta}{\delta\bar\varphi_{a}^{ABCD}}+\omega_{a}^{ABCD}\frac{\delta}{\delta\omega_{b}^{ABEF}}- \bar\omega_{b}^{ABEF}\frac{\delta}{\delta\bar\omega_{a}^{ABCD}}
\right.\nonumber\\
&-&\left. M_{cb}^{ABEF}\frac{\delta}{\delta M_{cCD}^{aAB}}- U_{cb}^{ABEF}\frac{\delta}{\delta 
U_{cCD}^{aAB}}+ N_{ca}^{ABCD}\frac{\delta}{\delta N_{cEF}^{bAB}}+V_{ca}^{ABCD}\frac{\delta}{\delta V_{cEF}^{bAB}}\right.\nonumber\\
&+&\left.R_{cb}^{ABEF}\frac{\delta}{\delta 
R_{cCD}^{aAB}}+T_{cb}^{ABEF}\frac{\delta}{\delta 
T_{cCD}^{aAB}}\right].
\end{eqnarray}
\item The Landau gauge condition is    given by
\begin{eqnarray}
\frac{\delta \Sigma_{BLG}'}{\delta b^{AB}} =\nabla^2D^a\Gamma_{aAB}.
\end{eqnarray}
\item The antighost equation  of motion yields
\begin{eqnarray}
\frac{\delta \Sigma_{BLG}'}{\delta \bar c^{AB}} +D^a \frac{\delta \Sigma_{BLG}'}{\delta K_a^{AB}}=0.
\end{eqnarray}
\item  The linearly broken, local  equation of motion of $\varphi_{aCD}^{AB}$
\begin{eqnarray}
\frac{\delta \Sigma_{BLG}'}{\delta \bar \varphi_{aCD}^{AB}}+D_b\frac{\delta \Sigma_{BLG}'}{\delta M^{Ab}_{baCD}}+C_{AB}^{EF,LM}T_{bEF}^{aCD}\frac{\delta \Sigma_{BLG}'}{\delta K_b^{LM}}=\nabla^2 C_{AB,GH}^{EF}
\Gamma_b^{GH}V_{EF}^{ba CD}.
\end{eqnarray}
The local, linearly broken, equation of motion of $\omega_{aCD}^{AB}$
\begin{eqnarray}
\frac{\delta \Sigma_{BLG}'}{\delta \omega_{aCD}^{AB}}+D_b\frac{\delta \Sigma_{BLG}'}{\delta N^{Ab}_{baCD}}+C_{AB}^{EF,LM}\bar\omega_{ EF}^{aCD}\frac{\delta \Sigma_{BLG}'}{\delta b ^{LM}}=\nabla^2 C_{AB,GH}^{EF}
\Gamma_b^{GH}U_{EF}^{ba CD}.
\end{eqnarray}
\item  The exact ${\cal R}_{aCD}^{bEF}$ symmetry reads
\begin{eqnarray}
{\cal R}_{aCD}^{bEF} &=&\int d^3x  \left( \varphi^{AB}_{aCD}\frac{\delta}{\delta \omega_{bEF}^{AB}}
- \bar\omega^{bEF}_{AB}\frac{\delta}{\delta \bar\varphi_{AB}^{aCD}}+
 V^{AB}_{caCD}\frac{\delta}{\delta N_{cbEF}^{AB}}-
  U^{bEF}_{cAB}\frac{\delta}{\delta M_{cAB}^{aCD}}\right.\nonumber\\
&+&\left. T^{AB}_{caCD}\frac{\delta}{\delta R_{cbEF}^{AB}}
  \right).
\end{eqnarray}

\end{itemize}
If we turn to the quantum level, these symmetries can be used to characterize the most general 
(allowed) BRST invariant counter terms.

\section{Conclusion}
The M2-branes worldvolume theory  have the following continuous symmetries:
16 supersymmetries, $SO(8)$ R-symmetry, nontrivial gauge symmetry and conformal symmetry.
The  multiple M2-branes described by the BLG theory,  which is based on Lie 3-algebras imposing  totally antisymmetric triple product (or 3-commutator). Though the 
BLG theory possess conformal invariance, the mass deformed BLG theory 
is no more a conformal invariant theory and can get the dynamics through 
non-vanishing $\beta$-function.
 So, it is important to investigate the mass deformed BLG theory 
 in GZ framework to investigate the theory in non-perturbative regime. Based on such reasoning,  this is possible to get a dynamical meaning to the
Gribov parameter.

We have considered the BLG theory with mass term in  $\mathcal{N} =1 $ superspace
quantized in Landau gauge in GZ framework. To avoid the Gribov 
copies from the LG theory in IR regime,
 a suitable non-local horizon term  restricting theory to the first Gribov horizon, has been added to 
the effective action. This non-local horizon term has further been localized by introducing a
 suitable 
quartet of auxiliary fields.  Further,  the BRST symmetry of the
BLG theory in  GZ framework has been addressed, where the $\gamma$-dependent breaks the BRST invariance. The BRST broken $\gamma$-dependent 
terms are extended further with three pairs of sources to restore the BRST invariance.
The various Ward identities  for such model, which help    to make the theory renormalizable, have been demonstrated.
With the help of these sets of Ward identities, one can compute  easily 
the suitable counter terms to absorb the divergences.
We believe that the present observation will improve
our current understanding of the issue of  Gribov problem in the supersymmetric Chern-Simons theory with 
Lie 3-algebras.
It would also be interesting to convert the soft BRST breaking of GZ BLG model
 in to the linear breaking, which 
guarantees the renormalizability of the theory. Because   the Quantum Action Principle suggests that the 
linearly broken BRST symmetry can
be directly converted into a suitable set of useful Slavnov-Taylor identities.

\end{document}